\documentclass[a4paper,12pt]{article}

\usepackage{graphics}

\begin{document}

\begin{center}
{\Large\bf Spreading and annihilating particles: surprises from mean-field
theory}\\
\vspace*{\baselineskip}
Georg Foltin$^1$, Karin A. Dahmen$^2$ and Nadav M. Shnerb$^3$\\
{\it$^1$Institut f\"ur Theor\-etische
Physik IV,\\
Heinrich-Heine-Universit\"at D\"usseldorf,
40225 D\"ussel\-dorf, Germany;\\
$^2$Department of Physics, University of Illinois,\\
Urbana Champaign, IL 61801;\\
$^3$Department of Physics, Judea and Samaria College, Ariel 44837, Israel.}\\

\vspace*{\baselineskip}
April 2000
\end{center}

\begin{abstract}
We study the pairwise annihilation process $A+A\rightarrow$ inert
of a number of random walkers,
which originally are localized in a small region in space.
The size of the colony and the typical distance between particles increases
with time and, consequently, the reaction rate goes down. In the long time
limit the spatial density profile becomes scale invariant. The mean-field 
approximation of this scenario bears some surprises. It predicts an upper 
critical dimension
$d_c=2$, with logarithmic corrections at the critical dimension and
nontrivial scaling behavior for $d<2$. 
Based on an exact solution of the one dimensional
system we conjecture that the 
mean-field {\it exponents} are in fact correct even below the upper
critical dimension, down to $d=1$, while the corresponding scaling function
that describes the spatial density profile,
changes for $d<2$ due to the stochastic fluctuations.
\end{abstract}

\begin{flushleft}
\end{flushleft}

%%% Introduction %%%

Recently there has been much progress made in the understanding of the 
kinetics of diffusing and reacting particles. In particular,
the evolution of spatial patterns in chemical and biological systems
has been subject to intensive research for many years. The effects
of (chemical) reactions between diffusively spreading particles, or
of ecological competition on a spreading population of species  
(in the absence of convection) can often
be described by a reaction diffusion equation of the type
\begin{equation}
\label{rdiff}
\partial_t {\mathbf n} = {\mathbf D} \nabla^2 {\mathbf n} 
+ {\mathbf f}({\mathbf n})
\end{equation}
where ${\mathbf n}(\underline{x},t)$ is the vector of reactants,
${\mathbf D}$ is a matrix of diffusivities, and ${\mathbf f}({\mathbf n})$
describes the nonlinear reaction kinetics \cite{Nel98}.
Equation (\ref{rdiff}) is really a deterministic approximation
(`mean-field'-approximation) of the underlying stochastic process,
replacing the discrete and fluctuating occupation numbers of the particles
by a continuous (coarse grained) population density.
In this paper we show that for the specific case of a single species
of diffusing particles $A$ that annihilate each other upon contact
$A+A \rightarrow$ inert, and for a certain set of (localized) initial 
conditions, this mean-field approximation is surprizingly 
powerful: it yields the correct scaling exponents in all dimensions,
even down to $d \ge 1$, while for $d<2$ the stochastic fluctuations 
due to the discreteness of the process change only the corresponding scaling 
function. Generally, in contrast, mean field theory is expected to 
yield exact scaling exponents
only {\it above} the upper critical dimension (which is 2 in this case),
wheras below, the stochastic fluctuations lead to anomalous exponents.
Right at the critical dimension we usually find logarithmic corrections to the
mean-field behavior. 

For stochastically {\it homogeneous} initial conditions, for example,
the $A+A\rightarrow$ inert reaction
\cite{Smo17,Kan85,Pel86,Lus86,Doe88,Fri92,Dro93,Gry95,Gry96,San96,Bar99}  
(and the closely related reaction  $A+A\rightarrow A$,
where particles undergo a Brownian motion until 
they hit another particle and undergo coagulation rather than annihiliation,
which is described by the same mean-field equation as the $A+A\rightarrow$ 
inert process) the density
decays as $n\sim t^{-d/2}$ for dimensions below 
the critical dimension 
$d_c=2$, and as $n\sim 1/t$ for $d>2$.
The latter formula can be deduced from the
deterministic (``mean-field'') equation of motion for the 
coarse-grained spatial density
$n(\underline{x},t)$ 
\begin{equation}
\label{mf}
\partial_t n=\nabla^2n-n^2,
\end{equation}
where we have absorbed the diffusion constant and the reaction rate into
the time and length scale. 
The $\nabla^2$-term describes again the diffusion of the particles, whereas the
$n^2$-term accounts for the pairwise annihilation process. For a homogeneous
density the mean-field equation reduces to $\dot{n}=-n^2$ leading to
$n\sim 1/t$.

For a certain set of localized initial conditions, however,
there is an  exception to the general change in behavior below
the upper critical dimension, due to the simple
re\-nor\-mal\-iza\-tion-group properties of the $A+A\rightarrow$
inert-reaction \cite{Pel86}. In this case the mean-field
equation (\ref{mf}) itself predicts correct
exponents, the upper critical dimension and logarithmical corrections:
at the beginning the system shall be empty up to a finite, populated region.
From this region particles will spread into the `vacuum' due to the diffusive
motion of the
particles; and for asymptotically long times we find a scale invariant 
density profile.
We will calculate this profile within the mean-field approximation and compare
it with an exact
solution of the one-dimensional problem.

The system under consideration serves as a toy
model for systems displaying a diffusive spreading of reacting agents
from a localized region
into space, e.g. models for chemical reactions which are ignited at some point.
It is related to but should not to be confused with the much more complicated
directed percolation problem \cite{Bro57,Dom84,Car96}
where, starting from a localized seed, particles can split, react, die 
and move diffusively.

A one-dimensional lattice version of the $A+A\rightarrow$ inert-reaction with
a homogeneous initial state was solved exactly
\cite{Lus86,Gry95,Gry96,San96,Bar99,Mat98}
owing to the fact that the stochastic matrix which enters the master equation can be expressed
as a bilinear form of fermionic operators. The mean density was found to decay as $n\sim t^{-1/2}$.
In addition, correlation functions were derived \cite{Gry95,Gry96,San96},
confirming the picture of a critical, non equilibrium
system with anomalous exponents. For dimensions $d>1$, an exact RG-approach \cite{Pel86,Fri92,Dro93,Lee94} established
the upper critical dimension $d_c=2$ and the power law $n\sim t^{-d/2}$ for dimensions $d<2$.
Consequently, the typical distance $\ell$ between particles grows like the
diffusion length $\ell\sim\sqrt{t}$.
This can be understood from the recurrence property of random walks for
$d\le 2$: Within a time intervall $t$ a Brownian particle reacts with
particles at distances smaller than the diffusion length $\ell$,
leaving behind voids of linear size $\sim\ell$ \cite{Smo17,Kra94,Red97}.

%%% Body %%%

In order to solve the mean-field equation (\ref{mf}) we substitute
\begin{equation}
\label{ansatz}
n(\underline{x},t)=\frac{1}{t}\phi\left(\frac{\underline{x}}{\sqrt{t}},t\right),
\end{equation}
reflecting the naive dimensions of time, length and the density $n$. It should
be kept in mind, that this
substitution is without loss of generality.
Plugging (\ref{ansatz}) into (\ref{mf}) yields
($\underline{r}=\underline{x}/\sqrt{t}$ is the rescaled position)
\begin{equation}
\label{rescaled}
t\partial_t\phi(\underline{r},t)=\nabla^2_r\phi+\frac{1}{2}\nabla_r\cdot
\left(\underline{r}\phi\right)+\left(1-d/2\right)\phi-\phi^2.
\end{equation}
Introducing a logarithmical
time variable $s=\log(t/t_0),$ where $t_0$ is an arbitrary time scale,
we arrive at
\begin{equation}
\label{rescaled2}
\partial_s\phi(\underline{r},s)=\nabla^2\phi+\frac{1}{2}\nabla\cdot
\left(\underline{r}\phi\right)+\left(1-d/2\right)\phi-\phi^2,
\end{equation}
which looks similar to the original equation of motion with, however, a deeply modified
propagator (linear part). We linearize equation (\ref{rescaled2})
in $\phi$ and split it into
\begin{equation}
\label{linearized}
\partial_s\phi=\Omega\phi+(1-d/2)\phi,
\end{equation}
where the operator $\Omega$, defined
through  $\Omega\phi=\nabla^2\phi+(1/2)\nabla\cdot(\underline{r}\phi)$ has zero
or negative eigenvalues. The unique groundstate (zero mode) of $\Omega$ is
$\phi_0=\exp(-r^2/4)$; the other eigenfunctions $\phi_{n_1,\ldots,n_d}=
\partial_1^{n_1}\ldots\partial_d^{n_d}\phi_0$ correspond to negative
eigenvalues $\lambda=-(n_1+\ldots+n_d)/2$
(using the notation $\partial_i\equiv\partial/\partial r_i$).

A critical dimension $d_c=2$ separates two qualitatively different sectors:
the linearized equation (\ref{linearized}) shows a growing mode for $1\le d<2$,
whereas all modes are exponentially damped in $s$ for $d>2$. 
For $1\le d<2$ and a nonzero localized initial condition, the unstable mode of $\phi$ will
grow until it saturates due to the presence of the nonlinear term. There is a unique,
stationary state $\phi^*$ of (\ref{linearized}),
given by $\nabla^2\phi^*+(1/2)\nabla\cdot(\underline{r}\phi^*)+(1-d/2)\phi^*-(\phi^*)^2=0$.
A closer inspection reveals that $\phi^*$ is stable, positive, has spherical
symmetry and decays faster than any power for $r\rightarrow\infty$.
Going back to the original coordinates via (\ref{ansatz}), we obtain the asymptotic time
dependence of the density $n(\underline{x},t)=t^{-1}\phi^*(\underline{x}/\sqrt{t})$.
Consequently, in $d<2$, the long time decay of the total number of particles,
for localized initial conditions is given by
\begin{equation}
\label{total}
N=\int d^dx\;n(\underline{x},t)\sim t^{d/2-1}, d<2.
\end{equation}
For $d=2$ we see in fact logarithmic terms.  The linearized equation
(\ref{linearized}) has a marginal mode (zero
mode) $\phi_0$. Therefore the time dependence of its amplitude $A$ is governed by the
nonlinear term: $\partial_sA\propto -A^2$ yielding $A\sim 1/s=1/\log(t/t_0)$.
Consequently, the total
number of particles is not constant as a naive extrapolation of (\ref{total}) would suggest, but
vanishes according to $N\sim 1/\log(t/t_0)$.
For dimensions above the critical dimension and a localized initial condition, $\phi$ behaves like
$\phi\sim\exp((1-d/2)s-r^2/4)$ as beeing the slowest mode of the
linearized equation (\ref{linearized}). The
nonlinearity turns out to be irrelevant since $\phi^2$ will vanish twice as fast as $\phi$.
Using the original coordinates, we have $n(\underline{x},t)\sim t^{-d/2}\exp(-x^2/(4t))$, i.e.
simple diffusion with a conserved total number of particles. In fact, for dimensions $d\le 2$ two particles meet with probability one, whereas for dimensions
$d>2$ two widely separated particles will miss each other and survive.

The remainder of this letter is devoted to an exact solution of the stochastic
one-dimensional system along the lines of
\cite{Lus86,Gry95,Gry96,San96,Bar99}. We represent an
occupied site $i=\ldots,-1,0,1,2,\ldots$ by an occupation number
$\tau_i=1$ and an empty site by $\tau_i=0$. We start with the
initial condition:
\begin{equation}
\label{initial}
\begin{array}{rccccccccccc}
\mbox{initial state:}&\ldots&0&0&1&1&\ldots&1&1&0&0&\ldots\\
\mbox{site:}&\ldots&-1&0&1&2&\ldots&L-1&L&\ldots,&&
\end{array}
\end{equation}
where we choose an even number of occupied sites $L$ in order to avoid
a trivial parity effect. Otherwise, one particle survives - we would see
simple diffusion of this particle on the long run (This parity effect is absent
in the mean-field approximation).

Now we write down a master equation for the probability $P(\{\tau_i\})$,
as done in \cite{Lus86}. At first, we define operators $\nu_i$ and $c_i$ via
\begin{eqnarray*}
\nu_iP(\ldots,\tau_i,\ldots)&=&\tau_iP(\ldots,\tau_i,\ldots)\\
c_iP(\ldots,\tau_i,\ldots)&=&P(\ldots,1-\tau_i,\ldots),
\end{eqnarray*}
a projection state $Q(\{\tau_i\})\equiv 1$ and an Ising-like operator $s_i=1-2\nu_i$
yielding -1 for an occupied site and +1 for an empty site.
The master equation reads $\partial_t P={\cal M}P$ with the master operator
\begin{equation}
 {\cal M}=\sum_i\left(c_ic_{i+1}-1\right)\left(\nu_i+\nu_{i+1}\right).
\end{equation}
It describes the exchange process $01\leftrightarrow 10$ and the annihilation process $11
\rightarrow 00$. 
The particular ratio between the exchange rate and the annihilation rate allows
for a quadratic representation in terms of fermionic operators but does not
constitute a special case since the ratio of both rates is not conserved under renormalization \cite{Pel86}. 
We introduce fermionic operators $a_i=\ldots\cdot s_{i-3}s_{i-2}s_{i-1}c_i\nu_i$ and
$a_i^\dagger=\ldots\cdot s_{i-3}s_{i-2}s_{i-1}c_i(1-\nu_i)$ and find for the master operator \cite{Lus86}
\begin{equation}
{\cal M}=\sum_ia^\dagger_{i+1}a_i+a^\dagger_ia_{i+1}-2a^\dagger_ia_i+2a_{i+1}a_i.
\end{equation}
Following \cite{Bar99} we are able to derive the expectation value $\left<\nu_i\nu_{i+1}\right>$:
\begin{equation}
\left<\nu_i\nu_{i+1}\right>(t)=\left<Q\right|\nu_i\nu_{i+1}\exp\left(t{\cal M}\right)
\left|P_0\right>,
\end{equation}
where $P_0$ is the initial state (\ref{initial}). We have $\left<Q\right|{\cal M}=0$ and 
$\left<Q\right|c_i=\left<Q\right|$ and therefore
\begin{eqnarray}
\lefteqn{\left<\nu_i\nu_{i+1}\right>(t)=\left<Q\right|a_{i+1}a_i\exp\left(t{\cal M}\right)
\left|P_0\right>}\nonumber\\
&=&\left<Q\right|\exp\left(-t{\cal M}\right)a_{i+1}\exp\left(t{\cal M}\right)\exp\left(-t{\cal M}\right)a_i\exp\left(t{\cal M}\right)
\left|P_0\right>\nonumber\\
&=&\left<Q\right|a_{i+1}(t)a_i(t)\left|P_0\right>,
\end{eqnarray}
where
$a_i(t)=\exp\left(-t{\cal M}\right)a_i\exp\left(t{\cal M}\right)$.
By differentiating this definition with respect to $t$ we find easily
$a_i(t)=\sum_jg_{i-j}(t)a_j$, where $g_{i-j}(t)=\exp(-2t)I_{i-j}(2t)$ is the
propagator of diffusion on the 1D-lattice ($I_k$ is a modified Bessel
function of order $k$). Using (\ref{initial}) and the definition of $a_i$ we
arrive at 
\begin{equation}
\label{secmom}
\left<\nu_i\nu_{i+1}\right>=\sum_{j,k=1}^Lg_{i+1-j}(t)g_{i-k}(t)\,\mbox{sign}
(k-j)(-1)^{j+k}.
\end{equation}
In order to obtain informations about the particle density we utilize the 
equation of motion for the moments $\left<\nu_i\right>\equiv n_i$ \cite{Sch95}:
\begin{equation}
\partial_tn_i=\left<Q\right|\nu_i{\cal M}\left|P_0\right>=n_{i+1}+n_{i-1}-2n_i
-2\left<\nu_i\nu_{i+1}\right>-2\left<\nu_i\nu_{i-1}\right>,
\end{equation}
where we have used $\nu_ic_i=c_i(1-\nu_i)$ and
$\left<Q\right|c_i=\left<Q\right|$.
Together with (\ref{secmom}) we get
\begin{eqnarray}
\partial_tn_i&=&n_{i+1}+n_{i-1}-2n_i-2e^{-4t}\sum_{j,k=1}^L\left(I_{i+1-j}(2t)-
I_{i-1-j}(2t)\right)I_{i-k}(2t)\nonumber\\
&&\mbox{}\times\mbox{sign}(k-j)(-1)^{k+j}
\end{eqnarray}
and with the help of $I_{k-1}(2t)-I_{k+1}(2t)=(k/t)I_k(2t)$ and after
 (anti-) symmetrizing  $j\leftrightarrow k$
\begin{equation}
\partial_tn_i=n_{i+1}+n_{i-1}-2n_i+\frac{e^{-4t}}{t}\sum_{j,k=1}^L|k-j|
I_{i-j}(2t)I_{i-k}(2t)(-1)^{j+k}.
\end{equation}
We are now in a position to perform the continuum limit. We replace the discrete position
$i$ by a continuous variable $x$ and the discrete Laplacian by its continuum version. Furthermore, we approximate the discrete propagator of diffusion $e^{-2t}I_{i-j}(2t)\approx 
(4\pi t)^{-1/2}\exp(-x^2/(4t))$, valid for $\sqrt{t}\gg L$ and $j= 1\ldots L$.  
We have $\sum_{j,k=1}^{L}|k-j|(-1)^{j+k}=-L$ for $L$ even \cite{footnote1}
and obtain \begin{equation}
%\label{exact}
\partial_tn(x,t)=\partial^2_x n(x,t)
-\frac{L}{4\pi t^2}\exp\left(-\frac{x^2}{2t}\right).
\end{equation}
As before, we find a scale-invariant solution with the ansatz $n(x,t)=\varphi
(x/\sqrt{t})/t$ ($r=x/\sqrt{t}$ is again the rescaled position):
\begin{equation}
\partial_r^2\varphi+\frac{1}{2}r\partial_r\varphi+\varphi-
\frac{L}{4\pi}\exp\left(-r^2/2\right)=0,
\end{equation}
yielding
\begin{equation}
\varphi(r)=\frac{L}{4\pi}\left(e^{-r^2/2}+\frac{r}{2}\,e^{-r^2/4}\int_0^rdy\,e^{-y^2/4}\right).
\end{equation}

\begin{figure}
\includegraphics{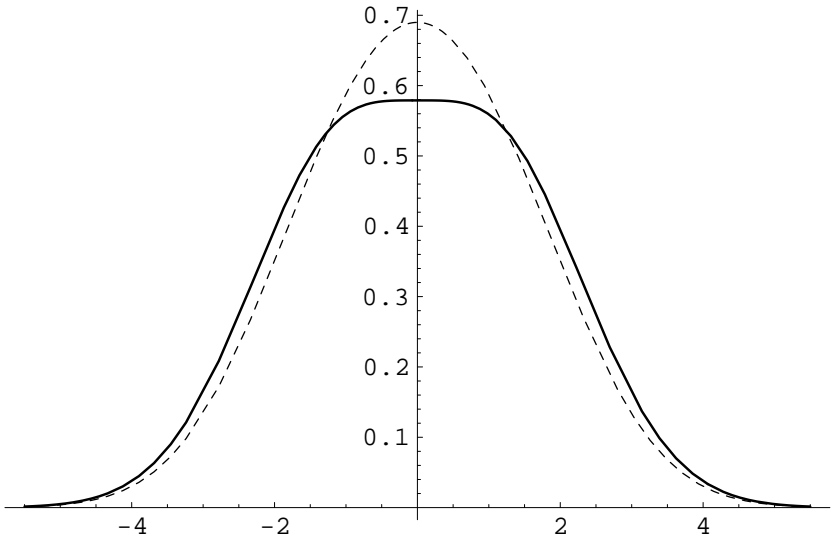}

\caption{\label{fig1}Exact rescaled density profile $\phi(r)$ (solid),
mean-field density profile (dashed - both for one spatial dimension).}
\end{figure}

Figure (\ref{fig1}) shows the exact profile $\varphi$ for the one dimensional problem and
in addition the corresponding mean-field profile, given by an appropriate numerical solution of
the nonlinear differential equation $\phi''+(r/2)\phi'+\phi-\phi^2=0$.  We have adjusted $L$ that
both curves have the same integral. Both, the stochastic $A+A\rightarrow$ inert
reaction in 1D with an initially localized configuration of particles and its
mean-field approximation do in fact share the same scaling exponents. We
find scale-invariant profiles $n(x,t)=\phi^*(x/\sqrt{t})/t$, and a total
number of particles decaying like $N\sim t^{-1/2}$.
The rescaled profiles $\phi$, however, show significant differences. The exact
profile has a {\it quartic} maximum $\phi(r)\propto 1-r^4/24$,
whereas the mean-field profile has a quadratic maximum
$\phi(r)\propto 1-0.31\times r^2$.
Correlations between different particles, which are neglected in the mean-field
approach, are responsible for the  difference.
For dimensions $d<2$ strong anticorrelations
between particle positions develop with time \cite{How97},
since two adjacent particles will recombine with high probability. Therefore
we have a central region with large voids and a relatively low density of particles and a number of particles, which have escaped the reaction zone.

The accuracy of the mean-field
approximation generally improves with
increasing spatial dimension, leading to the
conjecture, that the mean-field approximation predicts the correct scaling
exponents of the spreading colony also for dimensions $1<d\le 2$ (for dimensions
$d>2$ mean field is assumed to be correct anyway).

To conclude, we have studied the stochastic $A+A\rightarrow$ inert-reaction and 
have demonstrated that the mean-field approximation (\ref{mf}) performs
well for a particular initial condition.
If the particles are located initially in a small region in space from which
they start to spread out, the mean-field approximation
apparently predicts the correct
scaling exponents of the density profile due to the fact, that
the renormalization of the $A+A\rightarrow$ inert-reaction is quite primitive -
only the reaction
rate but not the propagator is getting renormalized \cite{Pel86,Lee94}.
This should be contrasted to the
homogeneous case, where mean-field (\ref{mf}) is wrong \cite{footnote2}
for $d\le 2$. Of course, in the homogeneous case, the mean-field equation
does not know anything about diffusion, whereas in the spatially inhomogeneous 
case the diffusion term is important.
Comparing the exact density profile for $d=1$ with its mean-field
approximation, we find significant differences and  show that stochastic
fluctuations are still relevant. 
It would certainly be interesting to see wether other systems with similar
RG-properties like the multi-species reaction scheme \cite{How96} display
analogous features.

%%% Acknowledgements %%%

We would like to thank U.C. T\"auber, D.R. Nelson, H.K. Janssen and R. Bausch
for helpful discussions. We acknowledge the hospitality of the Condensed
Matter Theory group at Harvard University, where some of this work was done.
GF was supported by the Deutsche Forschungsgemeinschaft through
Grant Fo 259/1.  KD gratefully acknowledges support from the
Society of Fellows of Harvard University.
This research was supported in part by the National Science Foundation
through Grant No. DMR97-14725 and by the Harvard Materials Research
Science and Engineering Center through Grant No. DMR98-09363.

%\bibliography{spread}

\begin{thebibliography}{10}
\bibitem{footnote2}
See, however, the more elaborated approach \cite{Smo17}, where the influx of
neighboring particles into a reference particle is calculated in a mean-field
like fashion, yielding correct results for the density decay.
\end{thebibliography}

\end{document}